\documentstyle[aps,prl,preprint]{revtex}
\begin{document}
\title
{Resonant two-magnon Raman scattering in antiferromagnetic insulators}
\author{Andrey V. Chubukov}
\address
{Department of Physics, University of Wisconsin-Madison,
1150 University ave., Madison, WI 53706\\
and P.L. Kapitza Institute for Physical Problems, Moscow, Russia}
\author{David M. Frenkel}
\address
{Texas Center for Superconductivity,\cite{byline}
University of Houston, Houston,
TX 77204-5932\\
and
Department of Physics, Science and Technology Center
for Superconductivity,\\
University of Illinois at Urbana-Champaign,
1110 West Green Street, Urbana, IL 61801}
\date{today}
\maketitle
\begin{abstract}
We propose a theory of two-magnon {\it resonant\/} Raman scattering
from antiferromagnetic insulators, which contains information both on
the magnetism and the carrier properties in the lighly doped phases.
We argue that the conventional theory does not work in the resonant
regime, in which the energy of the incident photon is close to the gap
between the conduction and valence bands.
We identify the diagram which gives the dominant contribution to
Raman intensity in this regime and show that it can explain the
unusual features in the two-magnon profile and in the two-magnon peak
intensity dependence on the incoming photon frequency.
\end{abstract}
\pacs{PACS: 75.10.Jm, 74.25.Ha, 78.30.-j}
\narrowtext

There is a widespread belief that strong electron-electron
correlations in the high-T$_c$ compounds may hold a clue to
the phenomenon of high-temperature
superconductivity~\cite{pinesanderson}.
One of the manifestations of these correlations is the fact
that the insulating parent compounds are antiferromagnets.
An important probe of antiferromagnetism is magnetic Raman
scattering~\cite{cottamlockwoodbook,singhreview}.
Its prominent signature in the underdoped high-$T_c$ materials 
is a strong peak observed at about $3000 cm^{-1}$. To first 
approximation, this peak can be attributed to inelastic scattering
from the two-magnon excitations~\cite{singhreview,fleury,canaligirvin}.

The traditional framework for understanding the two-magnon Raman scattering
in antiferromagnets has been an effective Hamiltonian for the interaction
of light with spin degrees of freedom known as the Loudon-Fleury
Hamiltonian\cite{loudonfleury},
$ H=\alpha\sum_{\langle ij\rangle} ({\bf\hat e}_i\cdot{\bf R}_{ij})
({\bf\hat e}_f\cdot{\bf R}_{ij}) {\bf S}_i \cdot {\bf S}_j,$
where ${\bf\hat e}_i$ and ${\bf\hat e}_f$ are the polarization vectors
of the in- and outgoing photons, $\alpha$ is the 
coupling constant, and ${\bf R}_{ij}$ is the vector
connecting two nearest neighbor sites $i$ and $j$.
Shastry and Shraiman\cite{shastryshraiman1}
have recently derived this Hamiltonian starting from the
large-$U$ Hubbard model.
Working in a localized basis, they performed a hopping expansion
controlled by $t/(U-\omega)$, where $t$ and $U$ are the nearest-neighbor
hopping and on-site Coulomb repulsion, and $\omega$ is of the order
of the photon frequencies.
The leading term in the expansion turned out to be the Loudon-Fleury
Hamiltonian.

This theory works well when the frequencies of the incoming
and outgoing photons are considerably smaller than the gap between the
conduction and valence bands, which is roughly $2 eV$.
The experimental reality in high-$T_c$ materials is such, however, that the
two-magnon scattering is measured mostly in the resonant regime, when the
frequencies of the ingoing and/or outgoing photons are close to the gap
value and the cross-sections vary strongly as the incident photon frequency
is varied\cite{oldjapaneseresonantdata,ranliu}.
Luckily, it is in this regime that the cross-sections sensitively depend 
not just on magnetic, but on the carrier properties as well,
and this makes understanding the data particularly important.

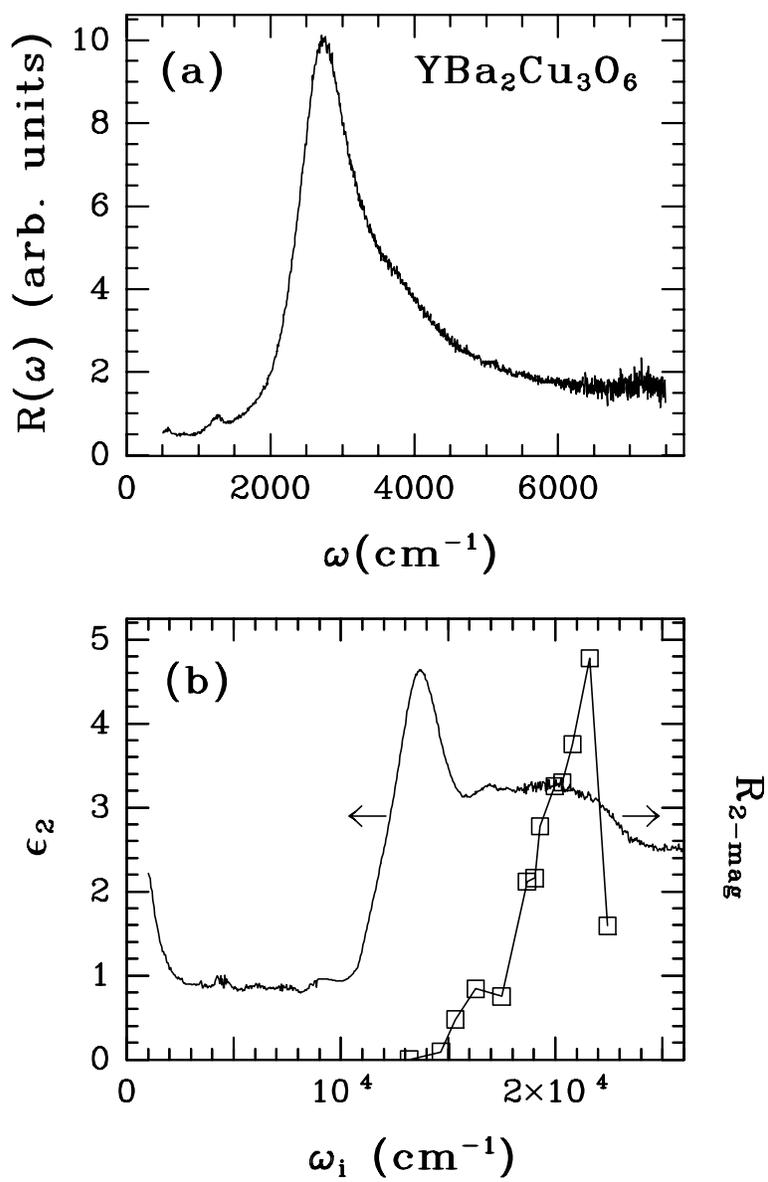
\begin{figure}
\caption{(a) A typical Raman cross-section 
in $YB_2Cu_3O_6$ as a function of transferred photon
frequency. A two-magnon peak is clearly seen. (b) 
The strength of the two-magnon peak as a function
of incoming photon frequency.
Also shown is the imaginary part of the dielectric constant.
Data courtesy of the authors of Ref.~\protect\onlinecite{ranliu}.}
\label{twom}
\end{figure}

The profile of the Raman cross-section as a function of the
{\it transferred\/} photon frequency 
and the  behavior of the two-magnon peak height as a function of the
{\it incident\/} photon frequency are shown in Fig.~\ref{twom}.

The key experimental features that require explanation
are: (a) the two-magnon peak is asymmetric, with the spectral weight
shifted to higher frequencies;
(b) the Loudon-Fleury Hamiltonian predicts no
scattering in the $A_{1g}$ configuration, whereas experimentally
the resonant $A_{1g}$ cross-section is about
half of that in the $B_{1g}$ geometry;
(c) there is only one peak in Fig.~\ref{twom}b --- ordinarily one might
expect two peaks, the so called ingoing and outgoing
resonances\cite{cardonareview,martinfalikovreview};
(d) a comparison with the dielectric constant shows that the strength
of the two-magnon Raman scattering in Fig.~\ref{twom}b is at its
maximum away from the band edge, in fact right at the upper end
of the features in the optical data that can be interpreted as the
particle-hole excitations between the lower and upper Hubbard bands.

In this paper we develop a diagrammatic approach to Raman scattering
valid in both nonresonant and resonant regimes.
For small frequencies, our results are identical to those
of Shastry and Shraiman.
However, for $|\omega -U| \sim {\cal O}(J)$, we find that the dominant
contribution to Raman scattering comes from a diagram which is
subleading in the nonresonant region.
We will argue that this diagram accounts for most of the
experimental features in Fig.~\ref{twom}.

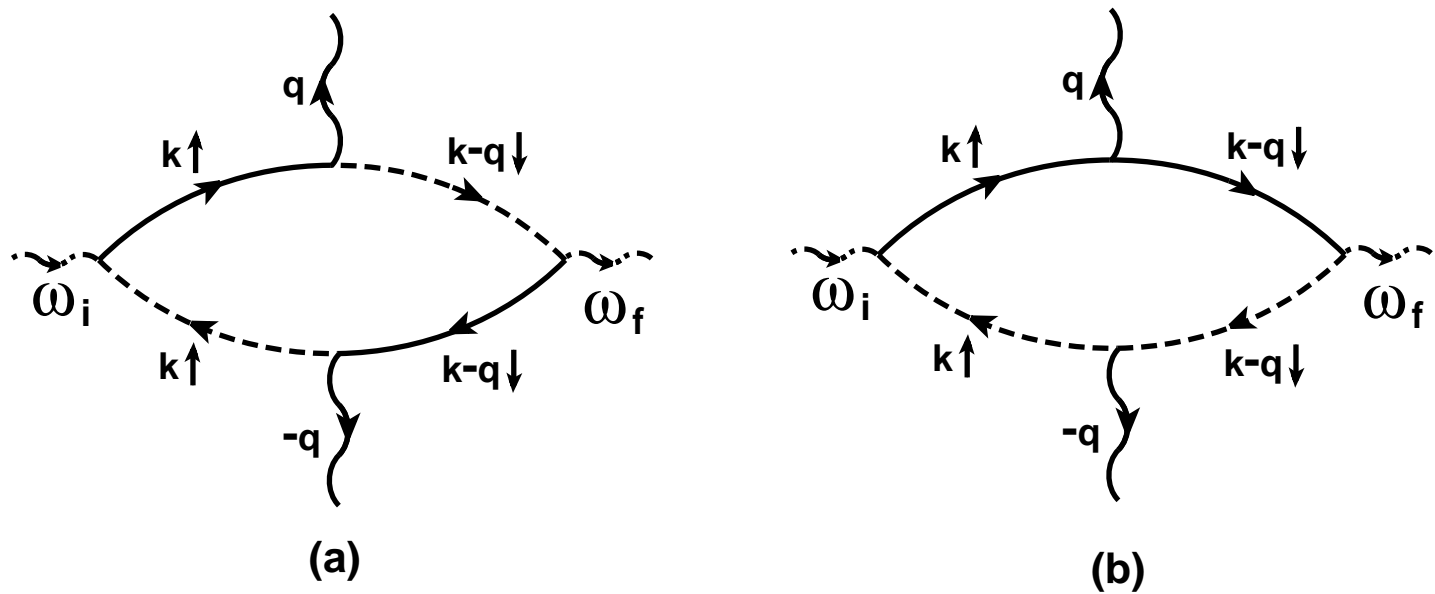
\begin{figure}
\caption{(a) A representative diagram which contributes
to the Loudon-Fleury Hamiltonian at small incident frequencies.
Each fermion can belong to either the valence (dashed line) or
conduction (solid line) band.
The emitted magnons are denoted by the solid wavy lines.
(b) The most singular diagram at resonance.} 
\label{nonresdiag}
\end{figure}

We start with the one-band Hubbard model with
$H= -t\sum_{\langle i,j\rangle} (c^\dagger_{i,\sigma} c_{j,\sigma} + h.c)    
+ U\sum_i n_{i\uparrow}n_{i\downarrow}$.
In the presence of the slowly varying vector potential ${\bf A}({\bf x},t)$
the Hubbard Hamiltonian gets transformed to 
\begin{equation}
H =  H^{{\bf A}={\bf 0}} -  \frac{e}{\hbar c} \sum {\bf j}_{q} \cdot
{\bf A}_{q} + O({\bf A}^2),
\label{n3}
\end{equation}
where
$j^{\alpha}_{q} = \sum_{k} \frac{\partial \epsilon_k}{\partial k_{\alpha}} 
c^{\dagger}_{k + q/2, \sigma}~c_{k -q/2, \sigma}$
is the current operator,
and $\epsilon_k=-2t(\cos k_x +\cos k_y)$ is the electron dispersion.
The resonant part of the scattering matrix element $M_R$ is obtained
from the term linear in $\bf A$ in the 2nd order of perturbation
theory\cite{cardonareview}.
In this process, a photon with energy $\omega_i$ and momentum which
can be safely set equal to zero, creates a virtual particle-hole
state of the fermionic system which can emit or absorb two spin-waves
with momenta $\bf k$ and $-{\bf k}$ before collapsing into an outgoing
photon with the energy $\omega_f$ (see Fig.~\ref{nonresdiag}).
Our primary goal is to calculate the dependence of this matrix
element on the incident photon frequency. 

We use the spin density wave (SDW) formalism\cite{schriefferwenzhang}
to describe the electronic state at half-filling and the excitations
around it. In the SDW formalism one introduces a long-range order in
${\bf S}_q = \sum_k ~c^{\dagger}_{k +q, \alpha} \sigma_{\alpha,\beta}
c_{k,\beta}$ with ${\bf q = Q} \equiv (\pi,\pi)$ and uses it to decouple
the Hubbard interaction term.
The diagonalization then yields two bands of electronic states (the conduction
and valence bands) with $E_k=\pm \sqrt{\epsilon_k^2 + \Delta^2}$, where the
$2\Delta \sim U$ in the strong coupling limit that is assumed throughout
this work.                  
In terms of the conduction and valence band quasiparticle operators 
$a_{k\sigma}$ and $b_{k\sigma}$, the current operator is interband
to leading order in $t/U$,
${\bf j}_{q=0} \Rightarrow
{\sum^{'}_k}{\partial \epsilon_k\over\partial {\bf k}}
(a^\dagger_{k\sigma} b_{k\sigma}+ b^\dagger_{k\sigma} a_{k\sigma})$.
We also need the magnon-fermion interaction.
Its derivation in the SDW formalism is straightforward, as the
magnons are described as collective modes in the transverse spin
channel~\cite{schriefferwenzhang}.
The answer is, for $S=1/2$,
 \begin{eqnarray}
H_{\rm el-mag} &=&  
{\sum_{k,q}}^{\prime}~(a^{\dagger}_{\alpha, k}a_{-\alpha,k+q}e^{\dagger}_{q}
~\Phi_{aa} (k,q) \nonumber \\
&+& a^{\dagger}_{\alpha, k} b_{-\alpha,k+q}e^{\dagger}_{q}
\Phi_{ab} (k,q) + (a \rightarrow b) + {\rm H.c.})
\label{tranham}
\end{eqnarray}
where $ e_q$ are the magnon operators, 
$\eta_q, {\overline\eta}_q = \frac{1}{\sqrt{2}}\left(\frac{1 \mp
\gamma_q}{1 \pm \gamma_q}\right)^{1/4}$, $\gamma_q = (\cos q_x + \cos q_y)/2$,
 and, to leading order in $t/U$, the vertex functions are given by 
\begin{eqnarray}
\Phi_{aa,bb} (k,q) &=& \left[\pm (\epsilon_k+
\epsilon_{k+q})\eta_q + (\epsilon_k-\epsilon_{k+q}){\overline\eta}_q
\right]; \nonumber \\
\Phi_{ab,ba} (k,q) &=& 2\Delta~\left[
\eta_q \mp
{\overline\eta}_q \right].
\label{vertices}
\end{eqnarray} 

In the situation when the photon frequencies are much smaller than
$\Delta$ all the energy denominators are of order $U$, and the
dominant diagrams for the Raman vertex are simply those with the
largest numerators.
It then follows from (\ref{vertices}) that one has to consider processes
in which the fermion changes bands while emitting a magnon.
A representative diagram is shown in Fig.~\ref{nonresdiag}a.
We collected all the leading order diagrams for the Raman matrix element
and obtained
\begin{equation}
M^{1}_R = \alpha\sum_{a=x,y} e_{ia}e^{*}_{fa}
\left[\cos q_a \left( \lambda^2_q + \mu^2_q \right)
- 2 \lambda_q \mu_q\right],
\end{equation}
where $\sqrt{2} \mu_q = {\bar \eta}_q + \eta_q,~\sqrt{2},
\lambda_q = {\bar \eta}_q - \eta_q$, and
$\alpha = 16t^2 \Delta/(4\Delta^2 - \omega^2)$.
This is exactly the expression which Shastry and Shraiman obtained
in their derivation of the Loudon-Fleury vertex for the Hubbard model.
Observe that within the Loudon-Fleury model,
 the scattering in the $A_{1g}$ geometry
(${\bf\hat e}_i ={{\bf\hat x} +{\bf\hat y}\over \sqrt{2}}$, 
 ${\bf\hat e}_f ={{\bf\hat x} +{\bf\hat y}\over \sqrt{2}}$),
vanishes, because in the $A_{1g}$ geometry the Loudon-Fleury and
Heisenberg Hamiltonians commute with each other.
On the contrary, in the $B_{1g}$ scattering geometry
(${\bf\hat e}_i ={{\bf\hat x} +{\bf\hat y}\over \sqrt{2}}$, 
${\bf\hat e}_f ={{\bf\hat x}-{\bf\hat y}\over \sqrt{2}}$)
the Raman vertex is finite,
$M^{B_1}_{R} \sim (\mu_q^2+\lambda_q^2)(\cos q_x -\cos q_y) =
{\cos q_x -\cos q_y\over \sqrt{1-\gamma_q^2}}$.
The profile of the two-magnon Raman scattering in the $B_{1g}$ geometry, 
including the final state interaction between magnons,
has been studied several times in the
literature~\cite{singhreview,canaligirvin}.
The two-magnon intensity has a narrow peak at the transferred frequency
$\omega \sim 3J$.

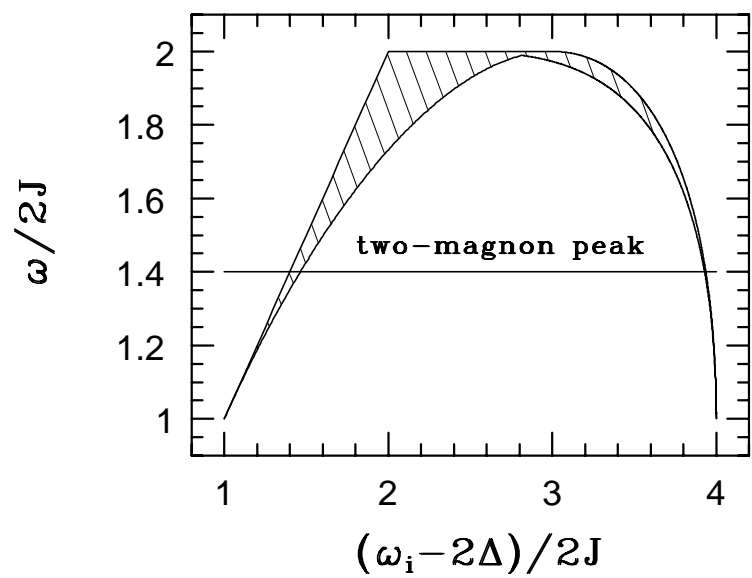
\begin{figure}
\caption{The triple resonance region (shaded) in the 
$(\omega_i,\omega)$ plane where $\omega = \omega_i -
\omega_f$. The horizontal line
corresponds to the position of the two-magnon peak which for definiteness
we chose to be at $\omega =2.8J$ which is the value one obtains in the 
$1/S$ expansion neglecting the renormalization of $J$~\protect\cite{CF}.}
\label{solution}
\end{figure}

A more careful treatment is, however, necessary in the resonant region,
when the incoming photon frequency is close to the gap value, and one
can no longer neglect the quasiparticle dispersion in the denominators.
In fact, most of the experiments on Raman scattering have been performed
in the frequency range where both $\omega_i$ and $\omega_f$ differ from $U$
only to order $J$.
In this situation, we found that the diagrams with the intraband
fermion-magnon vertices become dominant, since they contain more
resonant denominators.
We analyzed these diagrams and found that the most singular contribution
to the Raman vertex comes from the one in
Fig.~\ref{nonresdiag}b. 
The internal frequency integration in this diagram results in
  \begin{equation}
M^{(2)}_R = -8i{\sum_k}'
{
\left(
{\partial\epsilon_k\over\partial{\bf k}}
\cdot{\bf\hat e}_i
\right)\left(
{\partial\epsilon_{k-q}\over\partial{\bf k}}
\cdot{\bf\hat e}^*_f
\right) [ \mu_q\epsilon_{k-q}-\lambda_q\epsilon_k ]^2
\over (\omega_i-2E_k +i\delta)
(\omega_i-\Omega_q-E_k-E_{k-q} +i\delta)
(\omega_f-2E_{k-q}+i\delta)},
\label{dia5}
  \end{equation}
where $\Omega_q = 2J \sqrt{1 - \gamma^{2}_q}$ is the semiclassical 
spin-wave frequency.
A study of the integral shows that there is a region of $\omega_i$ and
$\omega_f$ where all three of the denominators in (\ref{dia5}) vanish
simultaneously, and the velocities
${\bf v}_k = \partial E_k /\partial {\bf k}$ and ${\bf v}_{k-q}$ are
antiparallel to each other (otherwise, the integral over $\bf k$ vanishes).
This phenomenon is known as a triple
resonance~\cite{cardonareview,martinfalikovreview}.
Via a combination of analytical and numerical techniques, we found that
for relevant $\omega_i$ the triple resonance in the Raman vertex occurs
only in a narrow range of the final photon energies $\omega_f$.
The region of triple resonances is shaded in Fig.~\ref{solution}.

It is important for our considerations that the triple resonance in the 
Raman vertex occurs only if both excited magnons are on the mass shell
(only then is the second denominator in (\ref{dia5}) a half-sum of the
other two).
This is true only for the diagram which does not contain final-state
magnon-magnon interactions.
On the other hand, for $S=1/2$, the dominant contribution to the 
conventional two-magnon peak at $\sim 3J$ comes from the diagrams
with magnon-magnon interactions~\cite{canaligirvin}.
In this situation the Raman spectrum $R(\omega)$ can be considered as
containing two {\it independent\/} peaks: one is due to the triple
resonance in $M_R$ in the shaded region in Fig.~\ref{solution}, which for
most of the experimentally measured $\omega_i$ is located close to $4J$, 
and the other, at transferred frequency of about $3J$, is due to the
magnon-magnon scattering. 
Without considering in detail the effects of the fermionic damping, which
smear the singularity in $M_R$, we cannot conclude which of the two peaks
is stronger.
The experiments indicate that the peak at $3J$ is stronger than that at $4J$,
and the enhancement of the Raman matrix element at larger transferred
frequencies is responsible for the observed asymmetric shoulder-like
behavior of the two-magnon profile.
Suppose we now fix  $\omega$ at the two-magnon peak frequency $3J$, as in
Fig.~\ref{twom}, and consider the variation of the peak amplitude as a
function of the incident photon frequency $\omega_i$.
Obviously, this amplitude will by itself have a maximum when the two peaks
in $R(\omega)$ merge, {\it i.e.\/}, when the $\omega=const$ line intersects
the region of triple resonances.
From Fig.~\ref{solution} we see that the intersection occurs
in a very narrow region of $\omega_i$ close to
$\omega^{max}_i = 2\Delta + 8J$, where the particle and hole are
excited near the {\it tops\/} of their respective bands~\cite{cc}.
We calculated Raman vertex in the vicinity of the intersection and found that
it diverges (in the absence of damping) as
  \begin{equation}
M_R \sim {\omega^{max}_i - 
\omega_i\over{(\omega_i-\omega_i^{res}+i\delta)^{3/2}}},
\label{reson}
\end{equation}
where $\omega^{res}_i$ can be well approximated by $\omega^{res}_i =
\omega^{max}_i - (\omega -2J)^2 /8J$.
The $3/2$ power of the denominator in (\ref{reson}) is due to triple
resonance, while the small factor in the numerator comes from the 
vanishing of the numerator in (\ref{dia5}) right at the top of the
band (i.e., at ${\bf k}={\bf 0}$).
In practice, the difference between $\omega^{res}_i$ and $\omega^{max}_i$
can be neglected, and Eq. (\ref{reson}) yields inverse square-root
singularity in $M_R$, which implies a linear singularity in the Raman
intensity, $R \sim |M_R|^2 \sim |\omega_i-\omega_i^{res}|^{-1}$.

Eq.~(\ref{reson}) is a key result.
In essence, we have found that the intensity of the two-magnon peak
increases by an inverse linear law as one approaches the upper edge
of the fermionic band.
We emphasize that the singularity at the top of the band exists,
due to the triple resonance effects, despite
the vanishing of the numerator in (\ref{dia5}) at this point.

We now discuss how these (and other) results are related to experiment.
We listed the key experimental features in the beginning of the paper.
Here we comment on each of them: (a) {\it asymmetry of the two-magnon
peak profile\/}: our theory predicts that for $\omega$ smaller than
$\omega^{res}_i$ the two-magnon peak profile should be asymmetric, with
the ``shoulder-like'' behavior at frequencies close to $\omega =4J$.
This is consistent with experimental observations.
In particular, the experimentally measured two-magnon profile in $Pr_2CuO_4$
was analysed~\cite{tomeno} and found to contain two peaks, a two-magnon peak
at $3000cm^{-1}$, and a smaller one at $4000cm^{-1}$, which is precisely
as expected from our calculations; 
(b) {\it selection rules\/}: 
the leading diagram in the resonance regime contributes to the scattering
in both $B_{1g}$ and $A_{1g}$ geometries.
The signals in both geometries have been observed in the experiments.
Recall that the Loudon-Fleury theory predicts scattering only in the
$B_{1g}$ geometry; (c) {\it a single peak\/}: our theory predicts a
{\it single\/} maximum in the two-magnon peak intensity measured as
a function of the incident photon frequency, while from the
``Loudon-Fleury" diagrams we might have expected two peaks,
one at $\omega_i = 2\Delta$, and the other at $\omega_f =2\Delta$
(the incoming and outgoing resonances)~\cite{cardonareview};
(d) {\it peak location and shape\/}: 
our theory predicts the  maximum of the two-magnon peak intensity 
measured as a function of $\omega_i$ right near the upper edge of
the quasiparticle fermionic band.
This is consistent with the measurements of the dielectric constant,
which show that the Raman scattering is strongest right at the upper
edge of those features in the optical data that can be interpreted
as particle-hole excitations between the lower and upper fermionic
bands.
We fitted the data on the peak intensity from Fig.~\ref{twom} by our
Eq.~(\ref{reson}) and found a satisfactory agreement with the 
predicted inverse linear dependence, except in the immediate vicinity
of the resonance, where the effects of fermionic damping
become relevant.

To summarize, we developed a diagrammatic approach to Raman
scattering in antiferromagnetic insulators which can be
used in both the resonant and nonresonant regimes.
We described for the first time the two-magnon Raman scattering in
the resonant regime, when the incident and final photon frequencies are
only ${\cal O}(J)$, apart from the gap between conduction and valence
bands.
This frequency range is relevant to recent experiments on undoped
high-$T_c$ compounds.
We identified the diagram which gives a dominant contribution to the
Raman vertex in this regime, and found the region in the
$(\omega_i, \omega_f)$ plane where the Raman vertex is strongly enhanced
due to triple resonance.
We demonstrated that the triple resonance, {\it combined with the SDW
dispersion relation for the carriers\/}, explains the unusual
experimental features in the two-magnon profile and in the two-magnon peak
intensity dependence on the incoming photon frequency.
In particular, our theory predicts the maximum of the two-magnon peak intensity 
right at the upper edge of the features in the optical data, as observed
in several materials~\cite{ranliu}.
This serves as a partial verification of the SDW picture for the
carriers, which, despite much theoretical work, has not been
well-established experimentally in these materials.

Beyond the scope of the present theory are the unexpectedly large 
width of the symmetric part of the two-magnon peak, which is probably
related to the magnon damping due to the interaction with
phonons~\cite{merlinpreprint1}, and the existence of a considerable Raman
signal $R(\omega)$ well above the maximum possible two-magnon energy
(i.e., $4J$)~\cite{girshprivate}, which may be related to chiral spin 
fluctuations~\cite{pasha}.

It is our pleasure to thank G.~Blumberg, C.~Canali, S.L. Cooper,
D.~Khveshchenko, M.V. Klein, R.~Liu, R.~Martin, R.~Merlin, H.~Monien, 
D.~Pines, S.~Sachdev, C.M. Varma, P.~Wiegmann and A.~Zawadowski for
useful discussions, and R.~Liu and S.L. Cooper for providing the
experimental data.
D.F. was supported by the Texas Center for Superconductivity at the University
of Houston and by the NSF Grant No. DMR 91-20000 through the Science and
Technology Center for Superconductivity at the University of Illinois.
Part of the work has been done while A.C. was at Yale University,
where he was supported by the NSF Grants No. DMR-8857228 and DMR-9224290.

\begin{references}
\bibitem{cc} There is a second crossing as $\omega_i \sim 2\Delta + 2J$, but
in this region only 
a small fraction of magnon directions, namely $q_x \approx q_y$ allow
for a triple resonance.
\end{references}
\end{document}